\documentclass{article}

\usepackage{arxiv}

\usepackage[utf8]{inputenc} 
\usepackage[T1]{fontenc}    
\usepackage{hyperref}       
\usepackage{url}            
\usepackage{booktabs}       
\usepackage{amsfonts}       
\usepackage{nicefrac}       
\usepackage{microtype} 
\usepackage{longtable}
\usepackage{graphicx}
\usepackage{doi}
\usepackage{subfigure} 
\usepackage{ragged2e}
\usepackage{pdflscape}
\usepackage{graphicx}
\usepackage{caption}
\usepackage{amsmath}
\usepackage{multirow}
\usepackage{tabularx}
\usepackage{amsfonts}
\usepackage{amssymb}
\usepackage{graphicx}
\usepackage{color}

\usepackage{amsmath}

\usepackage{amsfonts,amsmath}
\usepackage{amssymb, nccmath}
\usepackage{algorithm}
\usepackage{algorithmic}
\usepackage{amssymb, nccmath}
\usepackage{algorithm}
\usepackage{algorithmic}
\usepackage{dsfont}
\usepackage{booktabs}
\usepackage[utf8]{inputenc}
\usepackage{dsfont}
\usepackage{colortbl, hhline}
\usepackage{booktabs}
\usepackage{graphicx}
\usepackage{grffile}
\usepackage{grffile}
\usepackage{array} 
\usepackage{mathtools}
\usepackage{fontenc}
\usepackage{tikz}

\usepackage{afterpage}
\definecolor{yellow}{HTML}{E0B405}
\definecolor{timeLineProgressColor}{HTML}{E0B405}
\definecolor{currentPosition}{HTML}{A0C4A5}
\definecolor{gold}{HTML}{CA4F04}
\definecolor{beige}{rgb}{0.96, 0.96, 0.86}
\definecolor{frameBG}{HTML}{F5F5DC}
\definecolor{champagne}{rgb}{0.97, 0.91, 0.81}
\definecolor{apricot}{rgb}{0.98, 0.81, 0.69}
\definecolor{bronze}{rgb}{0.8, 0.5, 0.2}

\title{Workload Forecasting and Resource Management Models based on Machine Learning for Cloud Computing Environments}

\author{ \href{https://orcid.org/0000-0002-9689-6387}{\includegraphics[scale=0.06]{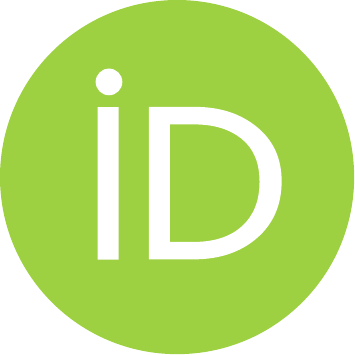}\hspace{1mm}Deepika Saxena}\thanks{The authors would like to thank National Institute of Technology Kurukshetra, India for financially supporting this research work.} \\
	Department of Computer Applications\\
    National Institute of Technology\\
	Kurukshetra, India \\
	\texttt{deepika\_6180096@nitkkr.ac.in} \\
	\And
	\href{https://orcid.org/0000-0002-8053-5050}{\includegraphics[scale=0.06]{orcid.pdf}\hspace{1mm}Ashutosh Kumar Singh} \\
		Department of Computer Applications\\
      National Institute of Technology\\
	Kurukshetra, India \\
	\texttt{ashutosh@nitkkr.ac.in} \\

}

\renewcommand{\shorttitle}{\textit{Predictive Resource Management for Cloud Environments}}

\hypersetup{
pdfsubject={q-bio.NC, q-bio.QM},
pdfauthor={David S.~Hippocampus, Elias D.~Striatum},
pdfkeywords={First keyword, Second keyword, More},
}

\begin{document}
\maketitle

\begin{abstract}
The workload prediction and resource allocation significantly play an inevitable role in production of an efficient cloud environment. The proactive estimation of future workload followed by decision of resource allocation have become a prior solution to handle other in-built challenges
like the under/over-loading of physical machines, resource wastage, Quality-of-Services (QoS) violations, load balancing, VM migration and many more. In this context, the paper presents a comprehensive survey of workload forecasting and predictive resource management models in cloud environment.  
 A conceptual framework for workload forecasting and resource management, categorization of existing machine learning based resources allocation techniques, and major challenges of inefficient distribution of physical resource distribution are 
discussed pertaining to cloud computing. Thereafter, a thorough survey of existing state-of-the-art contributions empowering machine learning based approaches in the field of cloud workload prediction and resource management are rendered. Finally, the paper explores and concludes various emerging challenges and future research directions concerning elastic resource management in cloud environment.
\end{abstract}


\section{Introduction}
Across the globe, advances in commercial computing and virtualization technologies have enabled the cost-effective realization of large-scale data centers, which execute large portion of internet applications. Cloud data centers offer numerous benefits 
including on-demand resources, elasticity, flexibility, mobility, and disaster
recovery \cite{buyya2018manifesto}, \cite{saxena2021op}, \cite{saxena2016dynamic}. Among these, elasticity is one of the most important characteristics of the cloud paradigm that allows an application to scale its resource
demands anytime \cite{dabbagh2015exploiting}, \cite{saxena2015highly}.
It has enabled the trend of renting of hardware, software and network resources rather than buying and managing computation resources. User can leverage complete computation infrastructure with an internet connection. It has wide range of applications like financial management, manufacturing, marketing, business management, academia, hospital management and many more \cite{gai2017resource}, \cite{saxena2018abstract}, \cite{saxena2021energy}. User is liberal to utilize high performance computation locally, can perform complex computation without bothering about its management constraints. Customers can enjoy benefits of cloud environment in the following ways:
\begin{itemize}
	\item Cloud environment provides ubiquitous computing service with ease and flexibility to all its customer over internet.
	\item Minimize start-up and operating cost investment, as it allows pay as per use service model. 
	\item Scalability feature of cloud data center enables to scale their applications by efficient utilization of IT resources, integration of distributed computing resources over the internet.
	\item Reduces the risk of resource allocation resulting from wrong estimation of workload or sudden power-off of resources and other management risks. 
	\item Tenants can collapse and expand their applications as per the requirement without bothering about capacity of resources. 
\end{itemize}
Consequently, massive amount of digital content is being produced through ever growing online activities such as shopping, social networking, email, and many others by the cloud clients. Cloud Service Provider (CSP) allocates computing instances called as virtual machines (VMs) to cater the resource demands of cloud clients.
Though cloud computing is increasingly used by public and private sector, the research on cloud computing is still at its developing
phase. Many existing challenges have not been fully addressed yet,
and new issues are rising from industrial applications. Some of the prevalent
research issues in cloud computing are:
\begin{itemize}
	\item{\textit{Automated Resources Provisioning}}\\ The main aim of a cloud service provider is to allocate and
	de-allocate resources efficiently to satisfy its SLA,  while minimizing its operational cost \cite{sharma2016multi}, \cite{amiri2018online}. It becomes difficult to determine how to map SLA such as Quality-of-Service (QoS) constraints to 
	low-level resource requirement such as CPU and memory
	requirements. Furthermore, to achieve high agility and respond to rapid demand fluctuations such as in flash crowd
	effect, decision of resource provisioning have to be taken online. Dynamic resource provisioning for internet applications provides solution for this. \\
	\item{\textit{Reduction of Energy and Power Consumption}}\\ Excessive power consumption due to wastage of resources at data center is another big challenge to deal with. The goal is not only to scale down energy cost in data centers, but also to
	meet government regulations and environmental standards. The solution for this is energy efficient resource scheduling, job scheduling, power shut down if in case not in use and proper utilization of CPU and other resources \cite{dabbagh2016energy}, \cite{saxena2015ewsa}. 
	
	\item{\textit{Data Leakage Detection}}\\ Security of user's data is most prevalant issue as customer's have to trust the third party for privacy and maintaining security of their confidential data \cite{singh2021cryptography}. Though service provider promise optimum security of data, still some malicious user actively hampers victim owner data. Moreover, VM migration should be safe as during migration target VM needs to traverse several architectural layers that should be trusted. Illegal authorization and authentication access raises more security overheads in cloud environment \cite{han2015using}. 

    	\item{\textit{Traffic Management}} \\It becomes serious issue to manage network traffic effectively so as to reduce congestion problem and other network related issues. So many reasons are there like, the density of links is much higher than
    	that in ISPs or enterprise networks, also the applications deployed on
    	data centers, such as MapReduce jobs,  have completely different traffic pattern \cite{saxenaa2020communication}.
    		\item{\textit{Task Scheduling }}\\ At cloud data centers, virtual machine allocation is done prior to the jobs arrival. Later, when applications from users arrive, they are divided into tasks that can independently execute on different VMs. It becomes a challenge to optimally distribute tasks on allocated VMs (resource provisioning) so that resources are maximum utilized and power consumption is minimum \cite{domanal2017hybrid}.
    	\item{\textit{Workload Prediction}} \\ Cloud computing promises elasticity, flexibility
    	and cost-effectiveness to satisfy SLA conditions. To realize these promises, service providers have to quickly plan and provision computing resources, so that the capacity of the supporting infrastructure can closely match
    	the needs of new demands. therefore, prior workload estimation is extremely important \cite{kim2020forecasting}, \cite{kumar2021self}. 
    		\item{\textit{Elasticity Management}} \\
    	Elasticity \cite{saxena2014review} is the ability of cloud system to 
    	adapt to workload changes by allocating and de-allocating resources in self adaptive manner,
    	such that at every instance of time the available resources match the current demand as closely as possible. It is directly related to speed of response and precision to map the tasks on VMs and VMs mapping to physical machines \cite{saxena2016dynamic}.
\end{itemize}

\section{Motivation}
The minimum upfront capital investment and maximum scalability features of cloud computing are fascinating to businesses, academia,  research and every public or private organizations. Evidently, smooth and progressive working of all these organizations depend on the services offered by the cloud data center. The daily life examples include social networking, e-governance, online shopping and others. In 2016, operational cost expenditure on public cloud Infrastructure as a Service of hardware and software
is forecast to reach \$38 B, growing to \$173 B in 2026.
\cite{Louis2016}. To meet the ever growing and dynamic demand of users, more and more virtual machines (VMs) are deployed on large number of servers and cooling devices are installed at data center that accounts for high power consumption \cite{sharma2016multi}. Infact, the operational cost due to power consumption is about 30\% more than resource purchase cost \cite{pedram2012energy}. The energy consumption distribution of cloud data center reported in \cite{rong2016optimizing} reveals that major portion of energy (40\%), is consumed by the number of active servers and cooling or refrigeration system employed at cloud data center. 
Due to these reasons data
center energy efficiency is now considered a chief concern for data center operators, ahead of the traditional considerations of availability and security. 
\par
\begin{figure}[!htbp]
    \centering
    \includegraphics{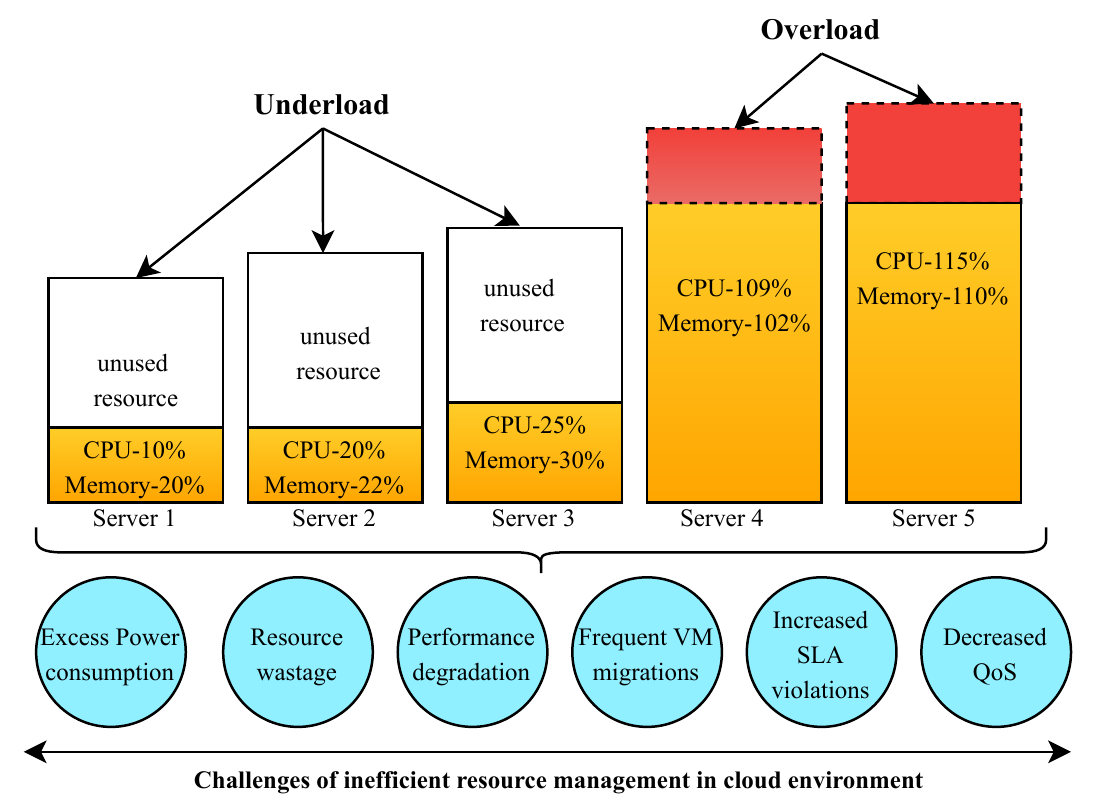}
    \caption{Typical resource utilization at cloud data center}
    \label{motivation}
\end{figure}

 Cloud users submit different type of workloads, requiring heterogeneous resource capacities with varying priorities and pricing policies associated with their respective Service Level Agreements (SLAs). The resource demand changes over time in an hour, day, week, month and years with respect to the type of workload and deadline of execution submitted by the user. Therefore, it becomes difficult to estimate the upcoming resource demands and decide resource distribution. It is observed that most of the time resources are either under utilized or overloaded  due to unbalanced VM placement leads to resource wastage and performance degradation \cite{dabbagh2016energy}. Even when running in the idle mode, servers consume a significant amount of energy.  Large savings can be made by turning off these servers. At the same time, these power saving techniques reduce system performance, pointing to a complex balance between energy savings and high performance.Typical model of resource capacity and utilization is depicted in Fig. \ref{motivation}. 
 
\par It clearly shows the
resource under-utilization and overload events during typical resource capacity of a data center. It also highlights the major challenges that arise due to inefficient distribution of physical resources at data center. To avoid
such scenarios the data centers are expected to have ideal resource capacity at any point of time such that it can fulfill all incoming requests without wasting resources. Again this can be achieved if resource
manager knows the upcoming workload in advance by accurate forecasting of
future demands. This emphasizes the need for automatic resource management techniques that enable
systems to auto-adapt according to the dynamic resource demands by using the existing resources more efficiently. \\

\textit{Our Contribution}: This paper specifically considers and presents a comprehensive survey of workload prediction and elastic resource management models. Furthermore, emerging challenges and future research directions in cloud resource management are rendered.
\section{Concepts and Categorization for Cloud Resource Management}

\begin{figure}[htbp]
    \centering
    \includegraphics[width=1.0\linewidth]{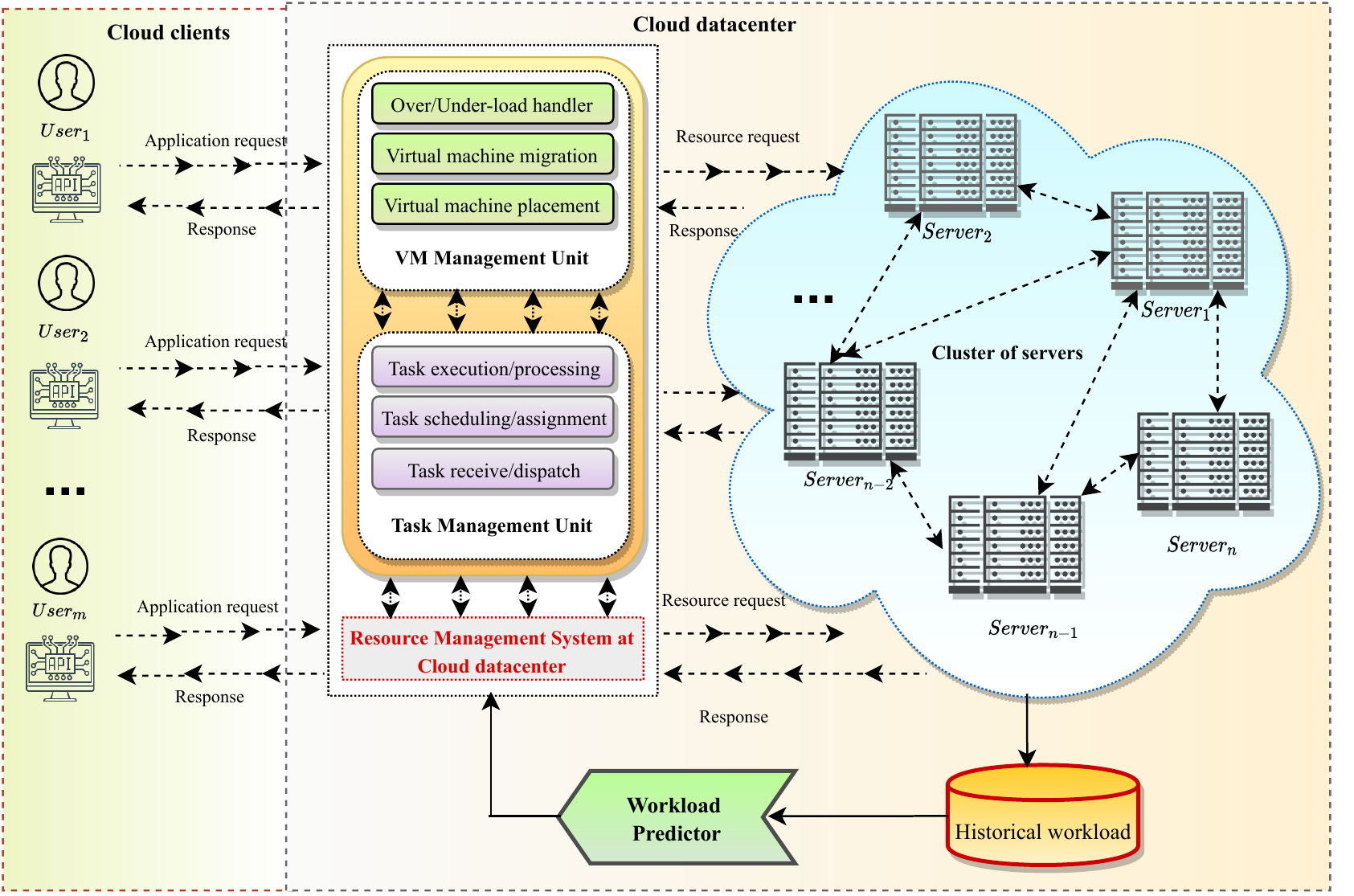}
    \caption{Conceptual framework for resource management in a cloud environment}
    \label{fig:my_label}
\end{figure}

\textit{Elastic Resource Management} is the 
process of allocation of computing, storage, networking 
resources to a set of applications, in order to jointly meet the performance objectives of the applications, the cloud service providers (CSPs) and the users of the cloud resources. The objectives of the CSPs is efficient and effective resource utilization within the constraints of Service Level Agreements (SLAs) with the cloud users. 
 Efficient resource utilization is typically achieved through virtualization technologies, which facilitate statistical multiplexing of resources across customers and applications. Fig. \ref{fig:my_label} shows conceptual framework for cloud resource management, where, $m$ users have requested 
different applications to be executed at the data center.
Each application has specific resource requirement that is to be satisfy by the data center. Resource Management System (RMS) is deployed at the data center to receive application requests from cloud clients and generate satisfactory responses for them, by allocating required capacity of resources (as per application demand) in the form of VMs. RMS consists of VM Management Unit (VMU) and Task Management Unit (TMU). VMU is responsible for VM scheduling and placement on physical machines (PMs), VM migration in case of over/under-load at PM. TMU works on applications, received from cloud clients/users, divide them into tasks, schedule and assign them on selected VMs for execution. The workload execution information is collected as a historical workload database which is used to train the workload predictor for estimation of future workload and resource utilization information. This predicted information is utilized for decision making of energy-efficient resource distribution and optimized load balancing.
\par
\begin{figure}[!htbp]
    \centering
    \includegraphics[width=1.0\linewidth]{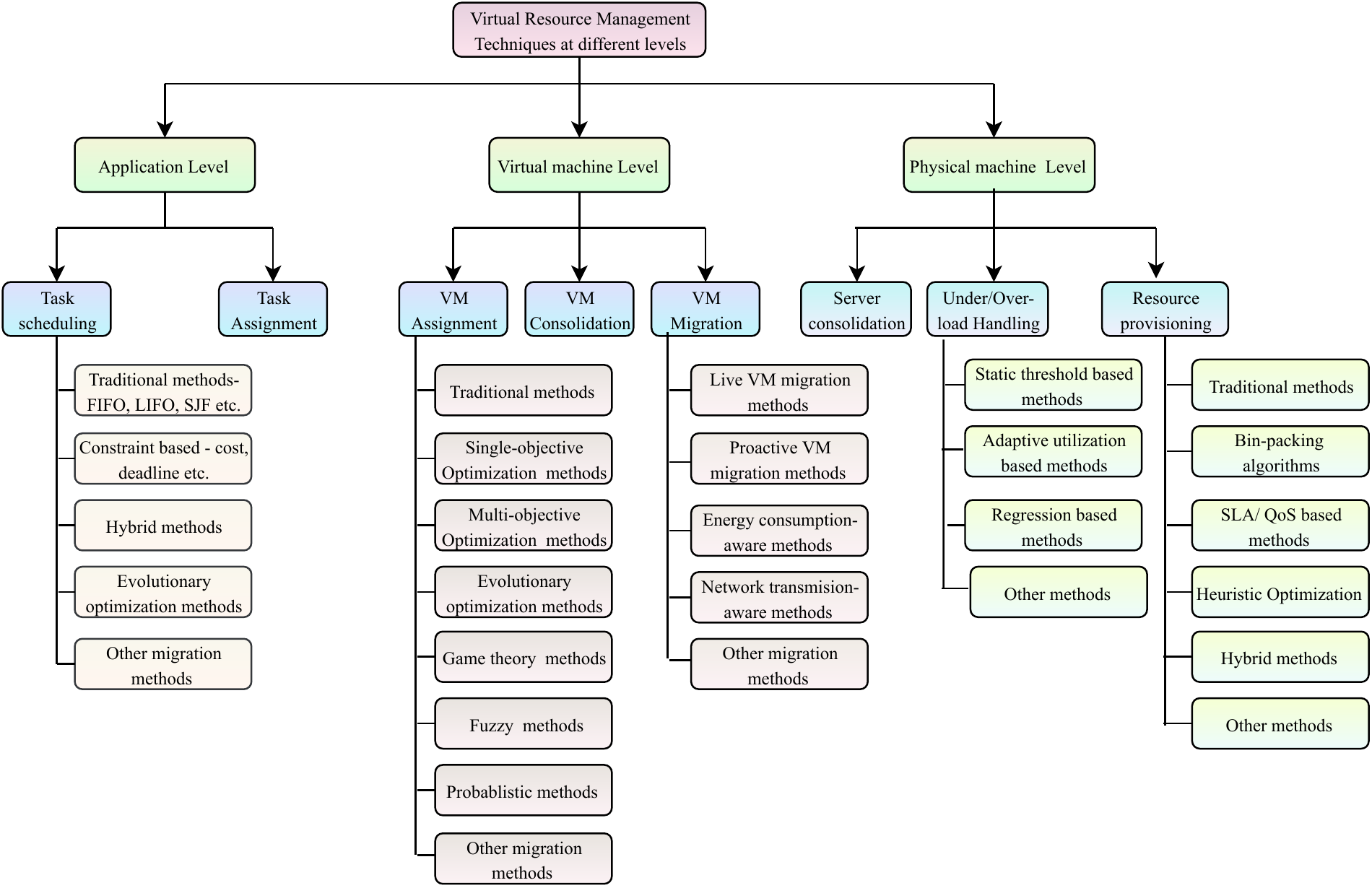}
    \caption{Categorization of Resource Management Techniques}
    \label{techniques}
\end{figure}

\textit{Virtual Resource Management Techniques}
 in cloud data center can be broadly categorized into workload management at application, virtual machine and physical machine levels  as shown in Fig. \ref{techniques}. At \textit{application level}, load management is required during task scheduling and assignment operations. Task scheduling means order of selection of task for execution while task assignment deals with selection of appropriate VM for task execution. Task scheduling and assignment can be performed by applying various methods shown in Fig. \ref{techniques}. The traditional task scheduling methods includes First-In-First-Out (FIFO), Last-In-First-Out (LIFO), Shortest Job First (SJF), Round-Robin (RR) scheduling etc. Also, the tasks may be scheduled on the basis of execution cost and deadline, or a multi-objective combination satisfying multiple objectives simultaneously with a priority. Furthermore, the multiple constraints based VM selection for a task execution is performed by applying traditional approaches like FIFO, LIFO, sorting VMs according to resource capacity, probabilistic or maximum likely-hood method, heuristic and evolutionary optimization algorithms.  \par 
 The workload management at \textit{Virtual machine level} can be divided into VM assignment or placement, VM migration and VM consolidation. \textit{VM assignment} is the process of deployment of VM on selected server while \textit{VM migration} is a process of shifting of VMs from over/under-loaded servers to a selected optimal server, and \textit{VM consolidation} is a technique to deploy VMs on minimum number of active servers to reduce power consumption and resource wastage. Fig. \ref{techniques} shows existing VM placement and migration methods. The VMs can be allocated to available servers in the cluster by utilizing conventional/traditional methods like, First-Fit, Best-Fit, Random-Fit, or Worst-Fit where VMs resource requirements are satisfied by the respective server without any contention of resources. The VMs can be placed according to single objective like, deadline of user task execution, maximum CPU, reduced power consumption, security, reliability or maximum availability etc. Also, the VMs can be allocated satisfying multiple objectives concurrently with priority or non-dominated sorting based pareto-front solution. To accomplish the same, commonly used approaches are evolutionary optimization (such as PSO, ACO, Firefly, GA, and hybrid optimization algorithms etc.), Fuzzy clustering, game theory based VM allocation, greedy approach, probabilistic method, and other optimization methods. 
 The VM allocation is further optimized by migrating VMs in case of resource contention by using live VM migration method i.e., hot VM migration and offline VM migration i.e., cold migration as per the requirement or VM movement decided by the load balancing unit. Additionally, VMs can be migrated while satisfying specific objectives including, energy-aware, network-aware, security-aware, or reliability-aware constraints etc.  
\par The basic operations involved in management at \textit{physical machine level} are server consolidation, handling over/under-load at server, server resource provisioning. The server consolidation deals with minimizing the number of active servers by applying energy-efficient VM placement techniques. The over/under-load handling includes detection of respective condition at PM and mitigate its effect by employing VM migration. Resource provisioning is the efficient distribution  of physical resources among VMs prior to arrival of actual demand from cloud user. Various under/over-load handling and resource provisioning techniques are mentioned in Fig. \ref{techniques}. The under/over-loading problem is tackled by migrating VMs from under/over-loaded servers to selected efficient server, where the destination server can be selected by applying different approaches such as first-fit, best-fit, random-fit, fuzzy approach based selection, or heuristic optimization etc. This problem can be confronted either reactively or proactively, where the former approach migrates VM after detection of overload, while the later approach estimates under/over-load prior to its occurrence and mitigate its effect by shifting VMs proactively. The physical resources of a server are provisioned using traditional methods, bin-packing algorithm and heuristic optimization methods while satisfying SLA and QoS constraints.  

\section{Workload prediction} 
\begin{figure}[!htbp]
	\centering
	\includegraphics[width=1.0\linewidth]{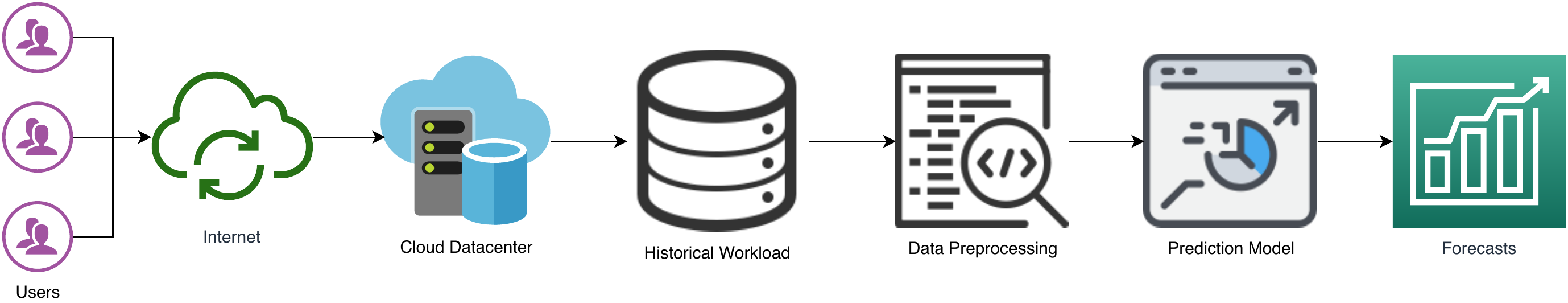}
	\caption{ General Workload Forecasting Model}
	\label{fig:predictionmodelworkflow}
\end{figure}

Cloud computing enables automatic resource scaling for every online transaction systems, which is one of the distinct key characteristics that distinguishes the cloud platform from the traditional computing models. However,
initializing a new virtual instance in a cloud is not instantaneous, cloud hosting platforms introduce some delay period while allocating hardware resources. Therefore, prediction based analysis of resource usage is the key to several
crucial system design and virtual machine deployment decisions such as, workload
management, system sizing, capacity planning and automatic
rule generation in the cloud. Fig. \ref{fig:predictionmodelworkflow} illustrates the general scenario of workload forecasting  at cloud data center. Millions of users sends request over internet at cloud datacener. The requests are processed there and all the request arrived during particular prediction interval are aggregated as historical data, which is later use to forecast workload of future. The historical data is collected and pre-processed in order to normalize it. Then normalized data is transferred to prediction system to forecast future workload. This predicted workload would give prior information of forth coming load and this allows effective time for preparation and arrangements of resource provisioning, decision to manage power on or off status. 

The general application forecast means to predict the future behavior of the
application in different ways such as the expected workload and the
performance. It becomes an important step to predict application in different dimensions to efficiently manage the resources in cloud.
As per the expected demands of application in future, the efficient
resource provisioning could be detected and the sufficient amount of resources could be planned/allocated to accomplish QoS parameters such as CPU utilization, response time, availability, reliability and security.\\

\textit{Different dimensions of workload forecasting}

The forecasting can be realized on virtual machines and physical
machines. At virtual machine level, it is necessary to maintain sufficient servers for virtual machine provisioning in order to balance upcoming workload at data center. The workload forecasting can be distributed into various dimensions depending upon the application as follows:
\begin{itemize}
	\item {\textit{Need for Prediction:}} The two very basic requirement that motivates prediction are resource management and application management. Resource management lays emphasis on efficient utilization of resources to raise fiscal gains, prevent SLA violation, to maintain consistent resource availability, prevent SLA terms and condition violation and avoid wastage of resources. Another requirement for accurate prediction is application management, which deals with load balancing of applications on virtual machines making decision related to VM migration, VM allocation, VM provisioning over physical machines.
	\item {\textit{Characteristics:}}  The various characteristics of workload prediction are accuracy of prediction outcomes, adaptability according to dynamic workload demands at different prediction intervals, proactive forecast in anticipation and effective utilization of historical data for future workload estimation.
	\item {\textit{Challenges:}} Commonly there are three challenges in workload forecasting. Time and space complexity needed in computing predicted workload should be reasonable in a way that its deployment is affordable. Data granularity is extremely important for workload forecasting. The coarse grained and long term data sampling causes the model to lose the
	 dynamic and adaptive behavior of the system. On the other hand, the fine-grained and short term prediction takes excess time and includes spurious details that are irrelevant and increase complexity to capture them.\\
	 Another big challenge is to decide the length of prediction, short prediction length leads to precise accuracy but have increased complexity. With larger prediction length, the number of data samples decreases and accuracy drops down. 
	
	\item {\textit{Evaluation Metrics:}}  The various metrics to evaluate prediction model are cost involved in predicting workload, success achieved in the form of accuracy of anticipating future demands, profit obtained in the form of intelligent resource management and error estimated during prediction.
\end{itemize}
\subsection{Recent state-of-the-art and Comparison }
This section exclusively dedicated to present a survey of recent cloud workload prediction models based on Neural Networks and Support Vector Machine as follows:

  \par {\textit{Evolutionary Neural Network}}\\
J.Kumar et.al, \cite{kumar2018workload} presented workload prediction approach for cloud data centers using neural network and self adaptive differential evolution (SaDE). In this work, request coming from different users are aggregated into specific time units as historical data.  Then, pre-processing of this data is done by normalizing it in the range of [0, 1]. Here, prediction model extracts patterns from actual workload and analysis of previous $n$ workload values to predict upcoming workload on data center at $n+1$ time instance. Then, self-adaptive differential evolution technique is applied to train the neural network. A biphase adaptive differential evolution (BaDE) learning algorithm trained neural network prediction model was proposed in \cite{kumar2020biphase}, which adopted dual adaptation viz., at level of crossover during exploitation process and mutation in exploration phase to improve the learning efficiency of neural network. This work outperformed \cite{kumar2018workload} in terms of prediction accuracy. Later in 2020, an auto-adaptive neural network was developed in \cite{saxena2020auto} where the network connection weights are optimized with a help of tri-adaptive differential evolution algorithm (TaDE). The adaptation is employed at crossover, mutation, and control parameters generation level which allows enhanced learning of evolutionary neural network. 
\par {\textit{Multi-input and Multi-output Neural Network}}\\
A multi-resource feed-forward evolutionary neural network based
 prediction model is proposed  in \cite{saxena2021proactive} by modifying the functionality of
an existing single input and single output (SISO) feed-forward neural network. There are sets of neural nodes instead of conventional nodes at input, hidden and output layers to receive multiple 
inputs and predict output based on multiple attributes. It employed combined classification and prediction operations 
during network weight connections learning process and concurrently classifies the predicted information of multiple attributes. It has improved efficiency in terms of prediction accuracy with lesser requirement of space and time complexity as compared to equivalent number of SISO neural networks required for prediction of same number of attributes. 
 \par {\textit{SVM and Feed-forward Neural Network}}\\
  Zhong et al. proposed a weighted wavelet support vector machine based host load prediction model in \cite{zhong2018load}. The approach combined the functionality of wavelet transform and support vector machine, and assigned weight to the sample, that reflected the importance of different samples and improved the accuracy of workload prediction. A Bayesian approach was proposed in \cite{kumaraswamy2017intelligent} for virtual machine workload prediction at data center. The method used resource demand forecasts to provision resources accordingly. A future workload prediction technique based on Back-propagation training on three-layered neural network was developed in \cite{prevost2011prediction} that produced workload prediction with acceptable accuracy on NASA HTTP web log traces for prediction interval upto 60 seconds. Cao et al. \cite{cao2018load} have applied historical data monitoring to predict the server load in  future by applying machine learning method like Random Forest. This scheme outperformed the time series analysis method in terms of accuracy for the workload prediction. 
  
  \par {\textit{Long Short-term Recurrent Neural Network (LSTM-RNN)}}\\
  The long short-term memory model in a recurrent neural network (LSTM-RNN) for fine-grained host load prediction was presented in \cite{song2018host} and \cite{kumar2018long}. Though the LSTM-RNN model learns long-term dependencies and produce high accuracy for prediction of server loads, it suffer from long computation time during training due to usage of Back-propagation algorithm between recurrent layers. 
  \par {\textit{Complex-Valued Neural Network}}\\
  In 2018  K.Qazi et.al \cite{qazi2018cloud}, proposed a workload prediction approach based on resource usage of host at data center. To train the prediction model, they applied neural network based on complex values that is more efficient than traditional real-valued based neural network. The complex value based neural network can further be developed as Quantum version of neural network, that would enable greater performance and speed benefits to the data center. 
  
  \par \textit{Ensemble Prediction Framework}\\
  CloudInsight workload prediction framework was proposed in \cite{kim2020forecasting}  that leverages an integrated potential of multiple machine learning predictors (for e.g., SVM, Neural network, Random Forest, Linear regression etc.) concurrently for accurate prediction of heterogeneous workload in cloud data center. Multi-class regression method is used to assign weights for different predictors in the ensemble predictor framework in real-time according to the current accuracy achieved by different predictors. The
ensemble prediction model is periodically optimized to tackle the sudden changes in the future workload. This prediction framework is inspired by the fact that an elastic resource management has inevitable dependency on workload predictors, which measures short-term and long-term deviations and fluctuations of heterogeneous cloud workloads. However, being pre-optimized for specific type of workload patterns learned during training via historical data, they are
insufficient to estimate the real-world cloud workloads whose patterns are unseen, may dynamically change over time, and may
be irregular. Consequently, these classical predictors often cause over-/under-provisioning of cloud resources.  The experimental simulation based performance evaluation reveals that CloudInsight has 13\%–27\% higher accuracy than classical machine learning based predictors. Also, it minimized period of under-/over-provisioning contributing to high cost optimization and low SLA violations.
  \par {\textit{Quantum Neural Network}}\\
  Singh et al. \cite{singh2021quantum} have proposed an Evolutionary Quantum Neural Network (EQNN) for prediction of heterogeneous kinds of cloud workloads including web, cluster, and high performance computing, etc. EQNN is an intelligent integration of specific principles of quantum computing and evolutionary algorithm trained neural network that utilizes C-NOT and rotation gates as an activation function within neural network. It is believed that quantum bit values (i.e., qubits) have better potential for exploration and exploitation than real numbers that allows intuitive learning of extracted patterns during training process. The pre-processed training input data is passed in the form of qubits generated with help of rotation gate and also the qubit values are generated for network connections instead of real-numbered weights.
  \\
  Table \ref{Table:workloadpredictionsummary} entails the summary of aforementioned machine learning based workload prediction approaches.
\begin{table}[htbp]
 \centering
 
  \caption{Summary of Machine Learning based Workload Prediction}
 	\begin{tabular}{ |  p{3.5cm} |  p{4cm} |  p{4cm} | p{3.5cm} |} 
 		\hline \textbf{Author, year }& \textbf{Approach} & \textbf{Strength}&\textbf{Weakness } \\ \hline \hline
 		 Prevost et al., 2011 \cite{prevost2011prediction} & Backpropagation trained neural network &  Expected load prediction per time unit & Less accuracy, longer convergence time\\ \hline
 	    
 	   Lu et al., 2016 \cite{lu2016rvlbpnn} & Random Variable Learning Rate Back-propagation Neural Network (RVLRBPNN) & outperformed HMM & works on single solution, slow convergence \\ \hline 
 	    Kumar et al., 2018 \cite{kumar2018workload} & Artificial Neural Network trained with SaDE evolutionary algorithm & improved prediction accuracy over Backpropagation & static approach for prediction \\ \hline 
 	   Kumar et al., 2018 \cite{kumar2018long} & Long  short-term  memory  model  in  a  recurrent  neural  network (LSTM-RNN) & better accuracy than ANN & long computation time \\ \hline 
 	  
 	  Zhong et al., 2018 \cite{zhong2018load} & Weighted wavelet SVM &  Variable load prediction accuracy & Requires hyper-parameters tuning with Longer convergence time\\ \hline

 	     Cao et al., 2018 \cite{cao2018load} & Random Forest ensemble approach &  Server load prediction with acceptable accuracy & Requires concisely pre-processed data, longer time \\ \hline
 	     Qazi et al., 2018 \cite{qazi2018towards} &  Multi-layered neural networks with multi-valued neurons (MLMVN) & high capability of learning  and better accuracy & higher time complexity than real-valued ANN \\ \hline 
 	  Saxena et al., 2020 \cite{saxena2020auto} & Neural network optimized by tri-adaptive differential evolution algorithm & Stable prediction accuracy & High time-complexity \\ \hline
 	  Kumar et al., 2020 \cite{kumar2020biphase} & Neural network optimized by bi-phase adaptive differential evolution algorithm & Faster convergence & Higher complexity than SaDE algorithm optimized neural network  \\ \hline
 	 Kim et al., 2020 \cite{kim2020forecasting} & Ensemble approach for adaptive learning of heterogeneous workloads & Real-time learning of cloud workload providing acceptable accuracy & Higher complexity than classical machine learning models  \\ \hline

 	      Saxena et al., 2021 \cite{saxena2021proactive} & Multi-input and Multi-output neural network optimized by modified differential evolution algorithm & Multiple related attributes are predicted concurrently & High complexity \\ \hline
 	  Singh et al., 2021 \cite{singh2021quantum} & Quantum based evolutionary neural network optimized by Self-balanced adaptive differential evolution algorithm & Higher and stable prediction accuracy & Higher complexity because of generation and processing of qubits \\ \hline
 	   
 	   \end{tabular}
 	   \label{Table:workloadpredictionsummary}
 	\end{table}  
\

 \section{Elastic Resource Management}
An elastic resource management in cloud environment comprises of various inter-related operations such as (\textit{i}) \textit{Task Assignment}, (\textit{ii}) \textit{VM Migration}, (\textit{iii}) \textit{Resource Prediction}, and (\textit{iv}) \textit{VM Allocation} as depicted in Fig. \ref{fig:rm}. All these operations interact with each other concurrently at common cloud platform to allow efficient distribution and management of available physical resources. Consider $m$ cloud users have submitted their requests for execution in cloud environment (Fig. \ref{fig:rm}), where Load Management Unit distributes different requests (i.e., applications divided into various tasks) on selected VMs by performing task assignment subject to the resource capacity constraints. The general functions performed by different interacting units with respect to the load management are as follows:
 \begin{figure}[!htbp]
	\centering
	\includegraphics[width=.84\linewidth]{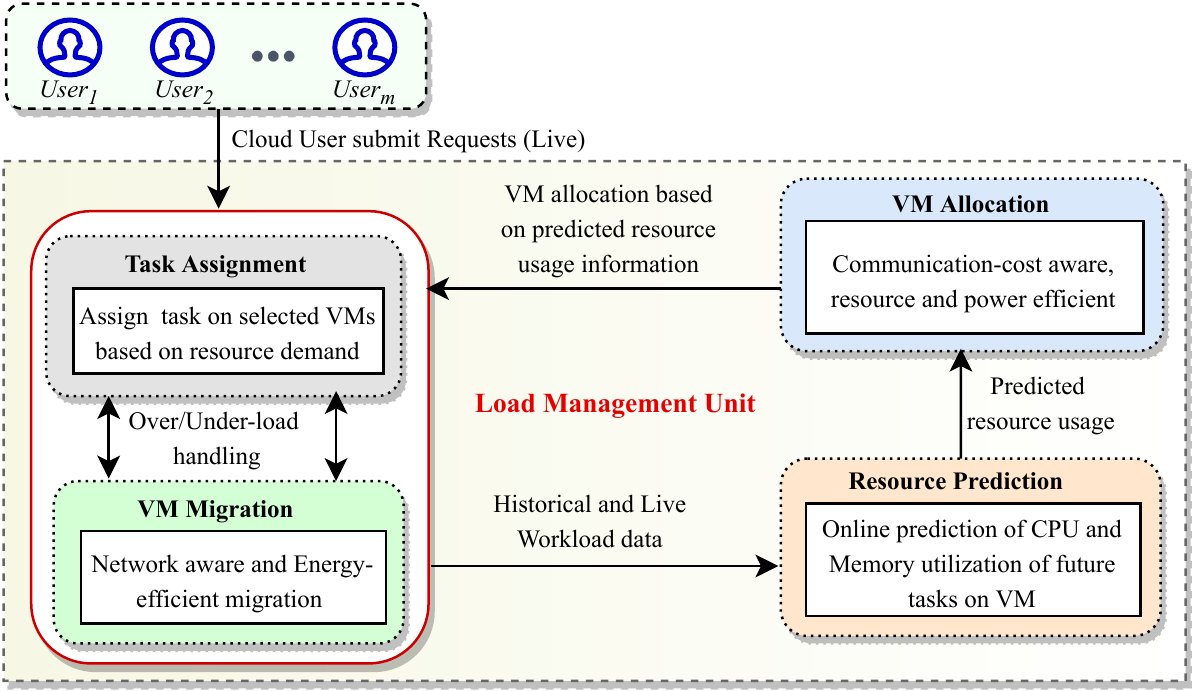}
	\caption{  Elastic Resource
Management Framework}
	\label{fig:rm}
\end{figure}
 
\begin{itemize}
  \item \textit{Task Assignment}: The tasks are assigned on the basis of their computational and storage requirements to the most appropriate VM and reduce wastage of resources as well. For instance, the tasks with small resource requirement are assigned to small size VM while large task to large size VM and so on, where size means resource capacity.
\item \textit{ Resource Prediction}: The expected demand of CPU and memory utilization of future tasks is determined based on the historical and current task execution by respective VM. Also, each VM prediction information is aggregated to detect any over/under-load on the server  beforehand for prior mitigation of its undesirable effects on performance of entire load management unit. 
\item	\textit{VM Allocation}: The VMs loaded with predicted resource usage are assigned to servers in energy efficient way such that power consumption and communication cost among inter-dependent VMs are minimum and resource utilization is maximum by applying suitable VM placement strategy.
\item	\textit{VM Migration}: To handle any over/under load (which severely degrades the performance of data center) during VM allocation, migration of VM is the best solution. The predicted resource usage information is utilized for proactive detection of any such situation and handle it by applying an appropriate energy and network traffic efficient VM migration strategy so as to minimize the chances of performance degradation.
\end{itemize}
An efficient VM Assignment and migration allows energy and network traffic optimization during load management which is further optimized by applying an appropriate resource prediction.
Hence, the purpose of integration of all these operations during elastic resource management is to conserve the optimization achieved at each step by applying various greedy, heuristic, fuzzy-approach, machine learning and evolutionary optimization methods and finally encapsulates the entire energy optimization to allow smooth and cost-efficient performance on cloud data center.

\subsection{Recent state-of-the-art and Comparison }

 \par \textit{Hierarchical SLA-driven resource management:}\\
 Goudarzi et al. presents SLA-driven resource management specifically for reduction of SLA-violation and electrical energy consumption including cooling devices and CPU utilization at cloud data center in \cite{goudarzi2015hierarchical}. The authors consider VM management is the problem of deploying VMs on servers and assigning resources to them satisfying the criteria of minimizing  overall operational cost and SLA terms violation subject to resource availability and QoS constraints. In this approach, a set of decentralized decision managers (power and cooling managers) generating constraints and works in cooperation to make hierarchical decision for VM management so as to reduce the operational cost of data centers. Moreover, they apply periodic optimization to assign new VMs and re-allocate active VMs to physical machines according to the expected workload during next epoch. The outcome of this work 40\% reduction in operational cost of data center and execution time is seven times lesser than centralized management structure for the same operations.
 
\par \textit{Euclidean distance based multi-objective energy efficient VM placement } \\
 Sharma et al. \cite{sharma2016multi} proposed an Euclidean distance based multi-objective energy efficient VM placement. This work includes energy consumption based on CPU utilization excluding cooling devices, multi-objective VM placement subject to minimum power consumption and maximum resource utilization, task scheduling and VM migration decreasing SLA violation. The authors proposed HGAPSO algorithm by 
  combining genetic algorithm (GA) and particle swarm optimization (PSO) to allocate VMs on servers to minimize resource wastage and SLA violation. GA helps in migration of VM from source server to target server but it is slow in convergence and PSO assist GA in selecting optimal target server by allowing VM placement from non-energy efficient to energy-efficient server. Task scheduling is done subject to deadline constraint of application execution, therefore reducing chances of SLA violation. When task get completed, VMs terminates and re-allocation VM starts migrating to energy-efficient physical machine. This work is compared with branch-and-bound based exact algorithm and improves resource utilization and reduces power consumption. 
  
  \par \textit{Exponential Weighted Moving Average based Resource allocation}\\
  Dynamic resource allocation and VM placement is presented in \cite{xiao2012dynamic} that assigns data center resources depending upon the application requirement and allows energy efficient resource utilization. This work contributes towards overloading avoidance and preventing resource wastage. Firstly, the authors have predicted the future needs of the resource by computing exponential weighted moving average (EWMA) using past behaviors of VMs. This predicted resource demand is later use to allocate VMs on servers accordingly. Moreover, the concept of 'skewness' to measure the unevenness or non-uniformity in resource usage of a server is introduced, which is minimize to enhance the overall resource utilization of server. To avoid overloading of servers, degree of overload is measured periodically, if threshold of resource usage reaches, VMs are migrated to less utilized server. To allow green computing, underutilized servers are shut down after migrating running VMs to energy efficient server. The outcome is load prediction with overall resource utilization is improved avoiding overloading situation as compared to without load prediction scenario. 
  
  \par \textit{Stochastic load balancing for VM management:}\\
  Yu et al. \cite{yu2016stochastic} presented stochastic load balancing to address problem of overloading of servers due to  multi-dimensional and dynamic workloads for VM management at data center.  Overloaded servers leads to performance degradation as each application executing on it consumes long time to response and frequent VM migrations. Firstly, expected resource demand in particular prediction interval is estimated and probabilistic distribution of prediction errors is padded to the predicted output in order to avoid SLA violations, under estimation and over estimation of resource demands occur due to errors during prediction.In addition, to handle improper VM migrations occur due to overloaded servers and inefficient resource utilization, hotspot (overloaded servers) are identified and a heuristic algorithm is proposed to decide VM migrations from hotspot to underutilized servers and fairly allocate VMs on available servers. This work outperforms deterministic load balancing schemes subject to handling overloaded servers, resource utilization and number of VM migrations. The limitation of the work is they do not consider power consumption or energy saving aspect while load balancing, security issues, network communication cost among various VMs of  a user, network bandwidth required during VM migrations.
 \par  \textit{ Online VM placement:}\\
   Online VM placement for raising cloud provider's revenue is proposed in \cite{zhao2015online}. The authors notify that SLA violations and VM migrations along with energy consumption degrades fiscal gain of service provider at data center. To resolve these issues, they present first-fit and harmonic algorithms for online VM placement without considering VM migration. Moreover, they proved that problem of maximizing overall revenue subject to VM migration is NP-Hard. They devised Least Reliable First (LRF, using least number of migrations) and Decreased Density Greedy (DDG heuristic algorithm, for fine-grained greedy migrations that increase cloud provider's revenue) to handle VM migration process. During experimental evaluation, it is observed that Harmonic algorithm has more potential to raise revenue for cloud service provider as compared to First-fit algorithm. DDG is feasible for high SLA penalty scenarios, and LRF is suitable to vice-versa situation. The limitation is they ignore communication cost among VMs, bandwidth requirement during VM migration. 
   
  \par  \textit{VM prediction and migration for overcommitted clouds:}\\
   Dabbagh et al. \cite{dabbagh2016energy} proposed energy efficient VM prediction and migration for over-committed clouds. Resource over-commitment means placement of VMs on servers with capacities excess to their actual requirement. This leads to unnecessary reservation of resources and extravagant power consumption. In addition, new problem arises, i.e. PM overloading and VM migration overhead. To combat these issues, resource predictor and overload predictor are deployed at each VM and PM respectively, that predicts future resource demands of a VM in anticipation and accordingly VMs are placed on PMs subject to the resource capacity and overloading constraints of PM. Overload predictor of all PMs are connected to cluster manager that decides VM migration process. For energy efficiency, the idea is to deploy new VM or VMs supposed to migrate on the already running PMs satisfying overload threshold constraints rather than to start new PM. The foremost limitation of this work is deployment of resource predictor at each VM and overload predictor at every server, as it leads to unnecessary overhead of computation that consume extra time and space at each physical machine.
   
  \par  \textit{Secure and Energy Aware Load Balancing (SEA-LB)}\\
   Recently, Singh et al. \cite{singh2019secure} presented secure and energy aware load balancing (SEA-LB) framework for VM placement, in which they introduced concept of security by minimizing side channel attack along with power saving and efficient resource utilization while allocating PMs to VMs. To provide security to each VM seems fine however, it reduces resource utilization. In reality, every VM is not carrying confidential data and therefore, SCA prevention is not the essential requirement of every VM while load balancing. Moreover, they ignored the role of VM migration, bandwidth utilization and network communication cost during balancing VMs on PMs.
   
\par   \textit{Multiobjective genetic algorithm:}\\ 
   Multiobjective genetic algorithm for resource prediction and allocation is proposed in \cite{tseng2017dynamic}, addressing CPU and memory utilization of VMs and PMs and overall energy consumption of data center. The GA algorithm predicts resource usage before allocating PM to VM. Then VM placement algorithm is applied to maximize resource utilization and minimize energy consumption. The limitation of this work is that SLA violations and VM migrations may increase due to prediction errors. 
  \par  \textit{Multiobjective ACO:}\\
   Gao et al. \cite{yu2016stochastic} proposed multi-objective ant colony optimization (ACO) for optimal VM placement and efficient power consumption. The goal is to optimally place VMs on available PMs so as to minimize resource wastage and power consumption. The drawback of ACO is that it depends on quantity of pheromones to search optimal solution in search space and it is unsuitable to recursively improve the resource utilization.
\par \textit{Profile guided VM placement}\\
Recently, Ding et al.  \cite{ding2019profile} proposed profile-guided three-phase framework for VM placement based on profiling of applications, task classification and application assignment. However, it ignored handling of VM migration during over/under-load situations. \par \textit{Traffic-aware VM Placement }\\
Meng et al.\cite{meng2010improving} focussed on network traffic minimization and studied the effect of network resources while optimizing VM migration on host machines. A two-tier approximation algorithm is proposed to solve traffic-aware VM placement problem that resulted into increased throughput and decreased communication cost. 
\par \textit{Secure and Multi-objective VM Placement}\\
A Whale Optimization Genetic Algorithm (WOGA) is proposed in \cite{saxena2021securevmp} for secure and multi-objective VM placement (i.e., SM-VMP) with multi-efficient VM migration. This work focussed on minimization of security threats, communication
cost, power consumption and improved resource utilization concurrently. A novel probability based encoding-decoding system was introduced that encodes VM allocation as a whale position
vector by considering heterogeneous VMs environment. The performance evaluation and comparison of this work with various existing multi-objective VMP including GA, Whale optimization, Non-dominated based GA (NSGA-II), PSO, Hybrid of GA and PSO reveals its superiority for efficient VM allocation. 
\par \textit{Online Prediction based Multi-objective Load Balancing}
\\ An energy-efficient and online resource prediction derived
multi-objective load-balancing (OP-MLB) framework is presented in \cite{saxena2021op} where an evolutionary neural-network based prediction unit is
developed and utilized to estimate the future
resource utilization of VMs and proactive over/under-load estimation on servers. A non-dominated sorting based genetic algorithm is applied for a multi-objective VM placement. Additionally, an energy and network-efficient VM migration algorithm is proposed for the minimization of power consumption within data center. It allows power saving by shutting down inactive servers and minimizing the number of active servers, reducing VM migrations and maximizing the resource utilization.
\par \textit{Proactive  VM autoscaling }\\
To meet dynamic demands of the future applications, an energy-efficient resource provisioning framework is developed in \cite{saxena2021proactive}. This framework addressed the challenges including resource wastage, degradation of performance and Quality-of-Service (QoS) by comparing the application's predicted resource requirement with resource capacity of VMs and consolidating entire load on the minimum number of servers. An online multi-resource feed-forward neural network (OM-FNN) is developed  to forecast the multiple resource demands and predicted VMs are placed on energy-efficient servers. This integrated approach optimized resource utilization and energy consumption. 
\par \textit{Communication Cost Aware Resource Efficient Load Balancing}\\
Saxena et al. \cite{saxenaa2020communication} have presented a 
 communication cost aware and resource efficient load balancing (CARELB) framework to reduce the communication cost among VMs within the data center, minimize power consumption and maximize resource utilization. This framework allocates VMs with high 
affinity and inter-dependency physically closer to  each other. 
A Particle Swarm Optimization and non-dominated sorting based Genetic Algorithm i.e. PSOGA algorithm is proposed for optimized placement of VMs, where VM allocations are encoded as particles as well as  chromosomes. The experimental analysis depicts that CARELB framework achieved stable optimization of power consumption and resource utilization during VM placement. 
\par \textit{Security embedded dynamic resource
allocation}\\
To reduce the security risks during VM placement and sharing physical resources with multiple users, a security embedded dynamic resource
allocation (SEDRA) model is proposed in \cite{saxena2020security}. This model employed an analyser for identifying co-located inter-VM relations and detect server with malicious
hypervisor to mitigate security breaches. A random tree based VM threat detector and workload predictor operates in the 
background to allow secure and resource efficient allocation of physical resources among VMs. SEDRA a protects against the security breaches due to known user VMs and substantially decreases co-residency threats due to unknown VMs. Futher, this model is enhanced in \cite{saxena2021osc} to provide security during load arrival at data center and detect any contention prior to its occurrence.
 \\
 Table \ref{table:RM} highlights the summarised comparison of related work discussed above subject to objective of the respective proposed work, utilized approach, strength and weakness of the work.   
\begin{table}[!htbp]
 \centering
  \caption{Comparative Summary }
 	\begin{tabular}{ |  p{3.5cm} |  p{4cm} |  p{4cm} | p{3.5cm} |}
 		\hline \textbf{Author, objective }& \textbf{Approach} & \textbf{Strength}&\textbf{Weakness } \\ \hline
 	Goudarzi et al., 2015 \cite{goudarzi2015hierarchical},	Energy efficient resource management & Hierarchical decision for VM placement & SLA violations and operational cost reduced & security and bandwidth were ignored.\\ \hline 
 	  Zhao et al., 2015 \cite{zhao2015online}, Online VM placement & Least Reliable First (LRF) and Decreased Density Greedy (DDG) algorithms were proposed & SLA violations and VM migrations reduced & ignored prior estimation of expected workload \\ \hline 
 	  Yu et al., 2016 \cite{yu2016stochastic}, stochastic load balancing & Prediction of over/under-loaded servers and VM migration & reduced number of active servers and improved resource utilization &  ignored security and communication cost between inter-dependent VMs  \\ \hline 
 	  Xiao et al., 2012 \cite{xiao2012dynamic}, Dynamic resource allocation and VM placement & Prediction of future resource demands by using EWMA & threshold based overload prediction & SLA violations increase due to prediction errors\\ \hline 
 	    Ding et al., 2019 \cite{ding2019profile}, profile-guided three-phase framework for VM placement & Profiling of applications, task classification and application assignment & improved resource utilization & ignorance of VM migration \\ \hline
 	    Nguyen et al., 2017 \cite{nguyen2017virtual}, VM consolidation & Prediction of multiple resource utilization to handle over/under-loaded servers & reduction in number of active servers & risk of SLA violations \\  \hline
 	   Wang et al., 2010 \cite{wang2010coordinating} & VM migration after overload detection & power-efficient VM management & delay in execution of user application \\ \hline 
 	 Beloglazov et al., 2010 \cite{beloglazov2010adaptive} & Threshold-value based overload detection & proactive handling of overload & may lead to unnecessary VM migration \\ \hline 
 	  Dabbagh et al., 2016 \cite{dabbagh2016energy} & Wiener Filter based overload prediction & Power efficient VM migration & ignored under-loaded server\\ \hline 
 	  	 Sharma et al., 2016 \cite{sharma2016multi}, Multi-objective VM placement & Hybrid of PSO and GA algorithm 	& improves resource utilization and reduces power consumption 
 	 & resource wastage  \\ \hline 
 	  Singh et al., 2019 \cite{singh2019secure}, Secure and Energy Aware Load Balancing (SEA-LB) & Multi-objective VM placement by applying NSGA-II & consideration of security during load balancing & VM migration not discussed \\ \hline 
 	   Gao et al., 2013 \cite{gao2013multi} multi-objective VM placement & Ant colony optimization (ACO) & optimal VM placement and efficient power consumption & unsuitable to recursively improve resource utilization \\ \hline 
 	   Tseng et al., 2017 \cite{tseng2017dynamic}, Multi-objective VM placement & GA based prediction for VM placement &  energy-efficient resource utilization &SLA violations and VM migrations \\ \hline 
 	  Meng et al., 2010 \cite{meng2010improving} & Two-tier approximation algorithm to provide network traffic aware Vm migration & increased throughput and decreased communication cost & ignored under-loaded server\\ \hline 
 	   Saxena et al., 2021 \cite{saxena2021securevmp}, Secure and multi-objective VM placement & Whale Optimization integrated with non-dominated sorting based Genetic Algorithm (WOGA) is proposed & Secure VM allocation subject to minimum power consumption, resource wastage, and communication cost is achieved & VM mapping to the level of multiple clusters of server is not considered \\ \hline 
 	 Saxena et al., 2021 \cite{saxena2021op}, online predictive resource management & Online VM resource-usage based multi-objective VM placement is proposed using evolutionary optimization technique & Energy-efficient VM placement with proactive handling of overload & inaccurate overload prediction may lead to unnecessary VM migrations \\ \hline 
 	   \end{tabular}
 	   \label{table:RM}
 	\end{table}  
 	
 \begin{table}[!htbp]
  \renewcommand\thetable{2}
 \centering
  \caption{Comparative Summary continued}
 	\begin{tabular}{ |  p{3.5cm} |  p{4cm} |  p{4cm} | p{3.5cm} |}
 		\hline \textbf{Author, objective }& \textbf{Approach} & \textbf{Strength}&\textbf{Weakness } \\ \hline
 	  Saxena et al., 2021 \cite{saxena2021proactive}, Proactive autoscaling of VMs & VM resource prediction and clustering is utilized to determine exact number and size of VMs required to execute future workload & Power efficient VM scaling and allocation & ignored under-loaded server\\ \hline 
 	  	 Saxena et al., 2020 \cite{saxenaa2020communication}, Communication-cost aware VM placement & Hybrid of PSO and NSGA-II algorithm (PSOGA) is developed	& improves resource utilization and reduces response-time 
 	 & Cluster-level communication-cost aware VM mapping is not considered  \\ \hline 
 	  Saxena et al., 2020 \cite{saxena2020security}, Security Embedded Dynamic Resource Allocation (SEDRA) & Workload prediction and proactive VM threat detection mechanisms are applied to achieve secure VM allocation& consideration of security during load balancing & VM migration is not discussed \\ \hline 
 	   Saxena et al., 2021 \cite{saxena2021osc},  Secure inter-VM communication model with network-efficient resource allocation & Proactive detection of network and resource congestion and unauthorized VM access based threat mitigation  & Allows secure VM inter-communication and avoids illegal access based VM threats & Over-/Under-loading of servers and VM migration concepts are not discussed during VM scheduling\\ \hline 
 	   
 	   \end{tabular}
 	   \label{table:RM}
 	\end{table}

 \section{Emerging Challenges and Future Research Directions}
 In the light of existing elastic resource management approaches, the following challenges are encapsulated for the enhancement and development of more promising and influential management of elastic resources in cloud environment:
 \begin{itemize}
 	\item \textit{Improving Workload forecasting accuracy}: The VM resource prediction and workload forecasting accuracy needs to be enhanced to avoid SLA violations and allow contention free execution of user applications. The accurate estimation of requirement of resources in future is inevitable in providing reliable and highly available cloud services during efficient resource management.  The existing approaches are static which works on historical workload only. 
 	\item \textit{Overloading and Performance Degradation}: During peak load arrival, the aggregate demand of resources of VMs exceeds the available resource capacity of the server leading to overloaded servers leading to performance degradation and SLA violations. For example, some VMs may crash, longer unavailability of resources and increased response time etc.
    \item \textit{Live VM migration and Downtime}: In order to manage the dynamic and random  requirement of excess resource capacities or handle overload, VMs migrate in real time from overloaded server to other server having sufficient resource capacity, leads to delayed execution. Another major issue is to decide which VM should migrate, when to migrate, and to which server it should migrate?  
 
		\item \textit{ Single resource consideration during load balancing}: Most of the existing elastic resource distribution strategies considers single resource i.e. CPU only, whereas in real world cloud resource management all the resources are significant as contention of either of the resources: Memory, Bandwidth, disk usage, processor may become bottleneck degrading overall performance of the system.
		\item \textit{Interaction and Co-operation among related operations }: Individual consideration of workload prediction, resource provisioning, VM placement and migration, though all of these operations serves common objective of energy-efficient resource utilization. Therefore, all the related resource management operations must be employed at uniform platform to allow interaction among them for development of a pragmatic resource management solution in cloud environment.
		\item \textit{Lack of security while resource allocation}: Since security of data is the primary concern of every user, the physical resource distribution among different users' VMs must consider protection of data as a major perspective along with resource capacity constraints.
     
 \end{itemize}
\par The following research directions are designed in order to fill the research gaps and deal with aforementioned challenges:
\begin{itemize}
    \item  An improved improved workload prediction method with high performance stability is needed to be developed. The predictor should learn dynamically in real-time by training and re-training itself with historical as well as real-time resource usage samples by various VMs and servers. Also, multiple resources should be predicted concurrently by learning utilization of different resources from the training data samples.  
    \item  An effective under/over-load prediction method is required that can proactively estimate the peak and down load conditions with high accuracy which would alert the proactive VM migration before actual occurrence of the under/over-load and minimize performance degradation. 
    
    \item To optimize multi-resource utilization including CPU, memory, bandwidth concurrently during load balancing for contention-free execution of users' applications while conserving energy and reducing resource wastage in cloud environment.
    \item  To integrate workload forecasting with an efficient load balancing approach or resource utilization model at uniform platform to enhance the efficiencv during resource management.
    \item To manage the network traffic while tackling resource congestion problem during resource distribution and network-efficient load balancing.
    \item  Development of secure resource distribution models for cloud data center that allows secure execution of user workload, secure inter-communication among inter-dependent VMs with alleviation of an unauthorised access. 
\end{itemize}

\bibliographystyle{IEEEtran}
\bibliography{mybibfile}

\end{document}